\documentclass[twocolumn,twoside,slac_two]{revtex4}
\usepackage{graphicx}
\usepackage{fancyhdr}
\begin{document}

\title{Collisionless Plasma Shocks \\ Field Generation and Particle Acceleration}

%

\author{Troels Haugb\o elle}
\affiliation{Niels Bohr Institute, Copenhagen University, 
Juliane Mariesvej 30, DK-2100 Copenhagen, Denmark}
\author{Christian Hededal}
\affiliation{Niels Bohr Institute, Copenhagen University, 
Juliane Mariesvej 30, DK-2100 Copenhagen, Denmark}
\author{Jacob Trier Frederiksen}
\affiliation{Stockholm Observatory,
Roslagstullsbacken 21, SE-106 91 Stockholm, Sweden}
\author{\AA ke Nordlund}
\affiliation{Niels Bohr Institute, Copenhagen University, 
Juliane Mariesvej 30, DK-2100 Copenhagen, Denmark}
\begin{abstract}
Gamma ray bursts are among the most energetic events in the known universe.
A highly relativistic fireball is ejected. In most cases the burst itself
is followed by an afterglow, emitted under deceleration as the fireball plunges
through the circum--stellar media. 
To interpret the observations of the afterglow emission,
two physical aspects need to be understood: 1) The origin and nature of the
magnetic field in the fireball and 2) the particle velocity distribution function
behind the shock. Both are necessary in existing afterglow models to account for
what is believed to be synchrotron radiation. To answer these questions, we
need to understand the microphysics at play in collisionless shocks. Using 3D
particle--in--cell simulations we can gain insight in the microphysical processes
that take place in such shocks. 
We discuss the results of such computer experiments. It is shown how a 
Weibel--like two--stream plasma instability is able to create a strong transverse
intermittent magnetic field and how this points to a connected mechanism for in situ
particle acceleration in the shock region.  
\end{abstract} 
\maketitle
\thispagestyle{fancy}
\section{Introduction}
Many compact relativistic objects have strong outflows of plasma, which emit non thermal radiation in internal
collisions (e.g. clumps in quasar, micro quasar and AGN jets, internal shocks in gamma ray bursts (GRBs) and when
colliding with the surrounding medium  (e.g. afterglows in GBRs, supernova remnants; terminal AGN shocks). The
non--thermal radiation is emitted in strongly collisionless shocks \citep[e.g.][]{2000ApJ..538L.125K, 2001ApJ..560L..49P}.
Despite their importance and universality collisionless shocks are poorly understood. The non--thermal radiation is
believed to be emitted in the shock by relativistic particles accelerated by strong electromagnetic fields.
Naturally this fact poses the questions which mechanism is responsible for the electromagnetic
field and how are the particles accelerated to the ultra--relativistic energies implied by observations.
In \citeyear{1999ApJ..526..697M} \citeauthor{1999ApJ..526..697M} suggested that the Weibel or two stream 
instability was responsible for creating a strong magnetic field in the shock interface.
It was recently confirmed numerically in computer experiments using particle--in--cell (PIC) codes.
[\citeauthor{bib:astro-ph/0303360} \citeyear{bib:astro-ph/0303360};
\citeauthor{2004ApJ..608L..13F} \citeyear{2004ApJ..608L..13F}; 
\citeauthor{2003ApJ..595..555N} \citeyear{2003ApJ..595..555N};
\citeauthor{2003ApJ..596L.121S} \citeyear{2003ApJ..596L.121S}]. 
\citeauthor{2004ApJ..608L..13F} \citeyear{2004ApJ..608L..13F}] showed that the magnetic field
reaches an energy content of a few per cent of the equipartition value.
Fermi acceleration has, so far, been used to explain the existence of a power law distributed non--thermal electron
population. It has been shown to occur in test particle Monte Carlo simulation under assumptions of the  structure
of the magnetic field; but as pointed out by \citeauthor{bib:Niemiec} (2004), to further the progress, self consistent
models taking into account the back reaction and the detailed microphysics has to be made. Recently,
\citeauthor{bib:baring}
(2004) showed that particle distribution functions (PDFs) inferred from GRB observations are in conflict with
those predicted by Fermi theory and diffusive shock acceleration.
In this proceeding we report on 3D  PIC models, also presented in 
\citep{2004ApJ..608L..13F,bib:hededal2004},
of counter streaming plasma shells as a description of the shock
interface in GRB afterglows.
\section{Numerical Setup}
To perform the numerical experiments we use a relativistic 3D PIC code as described in
\citeauthor{2004ApJ..608L..13F} (2004)
that works from first principles by solving the full Maxwell equations for the electromagnetic field and move the
particles according to the Lorentz force. We set up two counter streaming ion election populations in the rest frame of
the densest population. The density jump is 3. The inflow Lorentz boost of the less dense population is 
$\Gamma=3$ (Run I) and $\Gamma=15$ (Run II) in the two experiments, that we report on here.
The boundaries are periodic in the $x$-- and $y$
direction, transverse to the flow, and open in the $z$ direction parallel with the flow. 
The box sizes are $(x,y,z)=(200,200,800)$ and $(125,125,2000)$ respectively. The
electron skin depth is $5$ and $3.3$ grid zones respectively, and
we use a mass ratio of $\frac{m_i}{m_e}=16$ to resolve both the ion-- and the electron dynamics. 
We use 16 particles per cell and our boxes contained almost $10^9$ particles. 
The plasmas are initially unmagnetised and cold ($v_{th} \simeq 0.01c$).
\begin{figure*}[!th]
\begin{center}
\includegraphics[width=18cm]{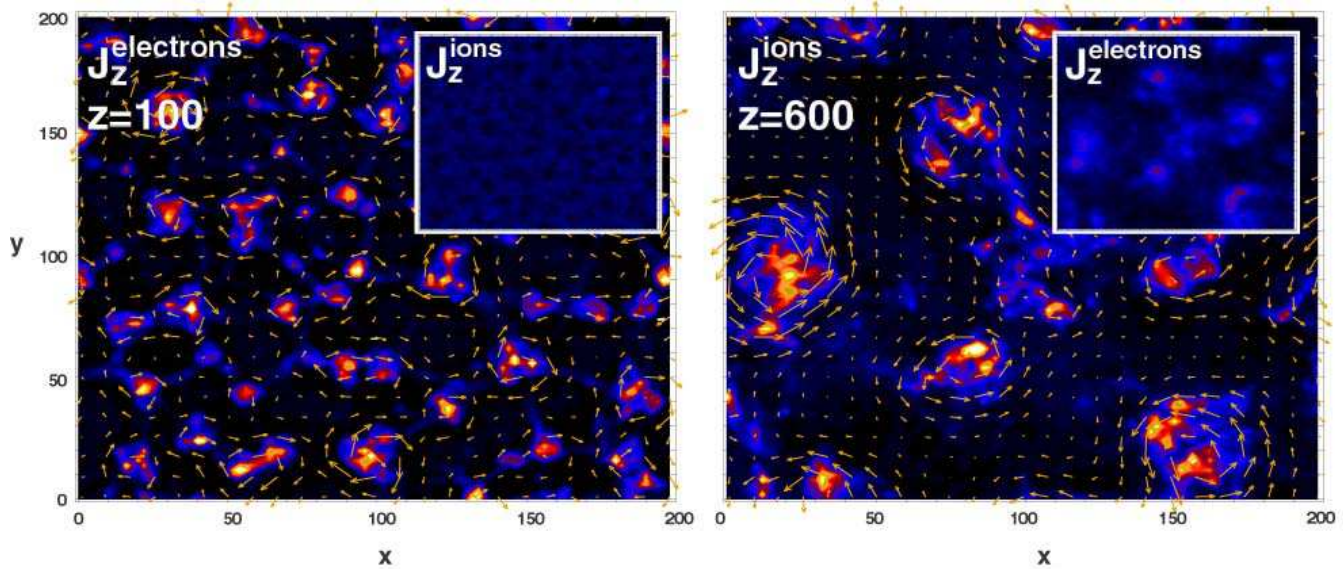}
\caption{The left hand side panel shows the longitudinal electron current
density through a transverse cut at $z=100$, with a small inset showing
the ion current in the same plane.  The right hand side panel shows
the ion current at $z=600$, with the small inset now instead showing
the electron current. The arrows represent the transverse magnetic field.
Both panels are from time $t = 1200$ in Run I.
\label{fig1}}
\end{center}
\end{figure*}
\section{Magnetic Field Generation}
In the computer experiments we see how the inter penetrating jets undergo the two--stream instability. First, the
electrons, being the lighter particles, collect into caustic surfaces and then current channels; in accordance
with the linear theory \citep{1999ApJ..526..697M}. Then they reach a non--linear saturation point and the channels
simply merge forming thicker and thicker channels.
When the magnetic field becomes strong enough the heavier ions are deflected into the magnetic voids between the
channels and start undergo the two--stream instability too. Since the ion instability is catalysed by the electron
instability the initial growth rate depends on the electron instability growth rate.
When the caustic surfaces of the ions collapse into current channels the electrons, being the lighter particles,
are heated and scattered by the magnetic field of the ion channels. 
Attracted by the electric potential of the ion channels the electrons start to Debye shield channels.
The Debye shielding quenches the electron channels, while at the same time it supports the ion channels. The large
random velocities of the electrons allow the ion channels to keep sustaining strong magnetic fields. This is
qualitatively different from the case of a pair plasma, where no shielding mechanism operates. The evolution is
illustrated in Fig.~\ref{fig1}. To the left we see the initial electron dominated phase,
while to the right --- further
downstream in the shock --- the ions dominate the dynamics forming dense ion channels; the more diffuse electrons
shielding the channels, and the resulting transverse magnetic field indicated with arrows.
The efficiency of conversion of the injected kinetic to magnetic energy, $\epsilon_B$, is around one per cent. The full
extent of the plasma dynamics operating in collisionless shocks is still not known,
but in \citeauthor{bib:hededal2004} (2004) it
was estimated that for a mass ratio $\frac{m_i}{m_e}$ of 16 and inflow Lorentz Boost $\Gamma=15$ (Run II)
the ion channels would
survive over 1500 ion skin depths. We speculate that with a more realistic mass ratios the Debye Shielding would be
more effective and therefore the ion channels would survive over even larger scales.
\begin{figure}[!b]
\begin{center}
\includegraphics[width=70mm]{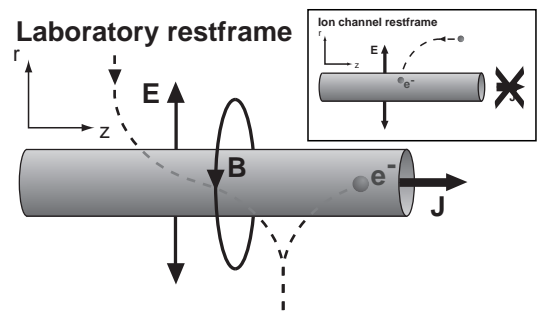}
\caption{A toy model of the acceleration mechanism: An ion channel surrounded by an
electric-- and magnetic field. Electrons in the vicinity of the current channel are
thus subject to a Lorentz force with both an electric and magnetic component
accelerating the electrons.
\label{fig2}}
\end{center}
\end{figure}
\section{Particle Acceleration}
\begin{figure}[!t]
\begin{center}
\includegraphics[width=70mm]{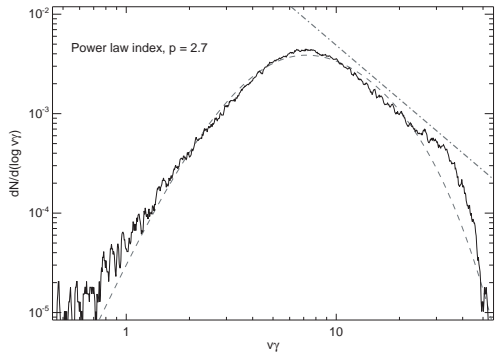}
\caption{The normalised particle distribution function
for the electrons in a slice around $z=1600$ downstream
of the shock in Run II. The dot--dashed line is a power
law fit to the non thermal high energy tail. The dashed curve
is a Lorentz boosted thermal electron population.
\label{fig3}}
\end{center}
\end{figure}

At distances less than a Debye length from the ion channels electrons are subject to a transverse acceleration
towards the ions since the electric field is not fully neutralised. There also exists a strong magnetic field and
in total this translates into a two component transverse oscillation and acceleration along the flow direction.
It was shown by \citeauthor{2004ApJ..608L..13F} (2004) that the ion channels are 
distributed according to a power law and later \citeauthor{2005ApJ..618L..75M}
(2005), showed that this is a consequence of the self similarity in the process of
merging ion channels. The acceleration a Debye shielding electron receives depends
on the size of the ion channel. As a direct consequence of the power law distributed
ion channels the electron PDF obtain a non--thermal energy tail as shown in
Fig.~\ref{fig3}. This is a local acceleration mechanism, that only depends on the
local electromagnetic field. Because the electrons are decelerated when moving away
from the channel; the principal energy losses are radiative
(e.g.~bremsstrahlung, synchrotron-- and jitter radiation).
It also implies that no high energy electrons are available for recursive acceleration
processes through this mechanism. We have confirmed this by an exhaustive search through
a representative dataset of $10^7$ particles. By tracking back and forth in time we
found only $\sim 5$ possible candidate particles, that had managed to escape
retaining their energy. It happened at places where the ion channel made sudden bends and
mergers and the electromagnetic fields were not well approximated as static fields.
None of them showed any sign of recursive acceleration. 
\section{Discussion}
In this contribution we have presented the results of self consistent PIC
computer experiments of collisionless shocks. We have shown how the two--stream
instability naturally generates a highly intermittent dominantly 
transverse magnetic field containing a few per cent of the equipartition energy.
The extent of the two stream instability in the case of an ion--electron plasma
is unknown; it clearly depends on the inflow Lorentz boost, the
density jump and the mass ratio. For $\Gamma=15$ a density jump of 3 and an 
electron--ion mass ratio 16 \citeauthor{bib:hededal2004} (2004)
estimated the instability to be sustained up to 1500 ion skin depths.
For realistic mass ratios this could be closer to $10^5$ ion depths. In the 
neighbourhood of the ion structures electrons are continuously accelerated
and decelerated. The mechanism is local and the power law distributed PDF
is created is situ. Hence the observed radiation may be tied directly to the
local conditions of the plasma. In this experiment with $\Gamma=15$ we found
a power law index of $p=2.7$. The mechanism does not rule out Fermi acceleration.
The lack of evidence in our numerical experiment for any recursive acceleration
processes can be due to the limited extent of the simulated region. It may, though, overcome
some of the difficulties pointed out by \citeauthor{bib:baring} (2004).
$\epsilon_B$, $\epsilon_e$ and $p$ should not be understood as free parameters.
To unravel the dependence on the outflow velocity and density jump a parameter study or
better theoretical understanding of the non--linear evolution is needed.
It is also clear that the non--thermal electron acceleration, the ion current 
channels and the magnetic field generation are beyond an MHD description and techniques 
respecting the full phase--space description of the plasma
is needed to further the understanding of collisionless shocks.

We thank the Danish Center for Scientific Computing for granting the computer
resources that made this work possible.


\begin{thebibliography}{9}   

\bibitem[{Baring \& Braby(2004)}]{bib:baring}
Baring, M.~G. \& Braby, M. 2004, ArXiv Astrophysics e-prints, astro-ph/0406025

\bibitem[{Frederiksen {et~al.}(2003)Frederiksen, Hededal, Haugb\o lle,
  \& Nordlund}]{bib:astro-ph/0303360}
  Frederiksen, J.~T., Hededal, C.~B., Haugb\o lle, T., \& Nordlund, {\AA}.
  2003, Proceedings of the 2002 Niels Bohr Summer Institute, ArXiv Astrophysics
  e-prints, astro-ph/0303360

\bibitem[{{Frederiksen} {et~al.}(2004){Frederiksen}, {Hededal}, {Haugb{\o}lle},
  \& {Nordlund}}]{2004ApJ..608L..13F}
  {Frederiksen}, J.~T., {Hededal}, C.~B., {Haugb{\o}lle}, T., \& {Nordlund},
  {\AA}. 2004, ApJ, 608, L13

\bibitem[Hededal et al.(2004)]{bib:hededal2004}
Hededal, C. B., Frederiksen, J. T., Haugb\o lle, T., \& Nordlund, {\AA}.
2004, ApJ, 617, L107

\bibitem[{{Kumar}(2000)}]{2000ApJ..538L.125K}
{Kumar}, P. 2000, ApJ, 538, L125

\bibitem[{{Medvedev} \& {Loeb}(1999)}]{1999ApJ..526..697M}
{Medvedev}, M.~V. \& {Loeb}, A. 1999, ApJ, 526, 697

\bibitem[{{Medvedev} {et~al.}(2005){Medvedev} {Fiore}, {Fonseca}, 
                 {Silva} \& {Mori}}]{2005ApJ..618L..75M}
 {Medvedev}, M.~V., {Fiore}, M., {Fonseca}, R.~A., Silva, L.~O.~\& Mori, W.~B.
 2005, ApJ, 618, L75

\bibitem[{Niemiec \& Ostrowski(2004)}]{bib:Niemiec}
  Niemiec, J. \& Ostrowski, M. 2004, ArXiv Astrophysics e-prints,
  astro-ph/0401397

\bibitem[{{Nishikawa} {et~al.}(2003){Nishikawa}, {Hardee}, {Richardson},
  {Preece}, {Sol}, \& {Fishman}}]{2003ApJ..595..555N}
  {Nishikawa}, K.-I., {Hardee}, P., {Richardson}, G., {Preece}, R., {Sol}, H., \&
  {Fishman}, G.~J. 2003, ApJ, 595, 555

\bibitem[{{Panaitescu} \& {Kumar}(2001)}]{2001ApJ..560L..49P}
  {Panaitescu}, A. \& {Kumar}, P. 2001, ApJ, 560, L49

\bibitem[{{Silva} {et~al.}(2003){Silva}, {Fonseca}, {Tonge}, {Dawson}, {Mori},
  \& {Medvedev}}]{2003ApJ..596L.121S}
  {Silva}, L.~O., {Fonseca}, R.~A., {Tonge}, J.~W., {Dawson}, J.~M., {Mori},
  W.~B., \& {Medvedev}, M.~V. 2003, ApJ, 596, L121
\end{thebibliography}
\end{document}